\documentclass[12pt]{article}
\pdfoutput=1
\usepackage[utf8]{inputenc}
\usepackage{amsmath}
\usepackage{amssymb}
\usepackage{units}
\usepackage{graphicx}
\usepackage{slashed}
\usepackage{array}
\usepackage[leftcaption,raggedright]{sidecap}
\usepackage{caption}
\usepackage[hidelinks]{hyperref}

\newcommand\pubnumber{NuPhys2023-Liz-Kneale}
\newcommand\pubdate{\today}

\def\napoli{for the Super-Kamiokande Collaboration}

\def\support{\footnote{
  University of Sheffield
}}

\def\Title#1{\begin{center} {\Large #1 } \end{center}}
\def\Author#1{\begin{center}{ \sc #1} \end{center}}
\def\Address#1{\begin{center}{ \it #1} \end{center}}

\newcommand\pubblock{\rightline{\begin{tabular}{l} \pubnumber\\
         \pubdate  \end{tabular}}}
\newenvironment{Abstract}{\begin{quotation}  }{\end{quotation}}
\newenvironment{Presented}{\begin{quotation} \begin{center} 
             PRESENTED AT\end{center}\bigskip 
      \begin{center}\begin{large}}{\end{large}\end{center} \end{quotation}}

\def\beq{\begin{equation}}
\def\eeq#1{\label{#1}\end{equation}}
\def\eeqn{\end{equation}}

\def\beqa{\begin{eqnarray}}
\def\eeqa#1{\label{#1}\end{eqnarray}}
\def\eeqan{\end{eqnarray}}

\let\bar=\overbar

\def\Dslash{\not{\hbox{\kern-4pt $D$}}}
\def\dslash{\not{\hbox{\kern-2pt $\del$}}}

\def\msb{{\bar{\ssstyle M \kern -1pt S}}}

\begin{document}
\begin{titlepage}
\pubblock

\vfill
\Title{Supernova spotting in Super-Kamiokande Gd}
\vfill
\Author{ Liz Kneale \support}
\Address{\napoli}
\vfill
\begin{Abstract}
Since the Super-Kamiokande (SK) neutrino experiment in Japan added gadolinium sulphate octahydrate (Gd) to the pure water in its detector, it has entered a new era in supernova neutrino detection. The addition of Gd makes it possible to tag neutrons from inverse beta decay (IBD) interactions of electron antineutrinos with free protons in the water. This has led to significant improvement in the experiment's sensitivity to low-energy electron antineutrinos, which is key in the observation of supernova neutrinos. Super-Kamiokande Gadolinium (SK-Gd) has the potential to make the first observation of the diffuse supernova neutrino background (DSNB) and pre-supernova (preSN) neutrinos, and can improve the pointing capability of SK in the event of a core-collapse supernova (CCSN) explosion. We present the latest results and prospects for supernova neutrino detection with SK-Gd.

\end{Abstract}
\vfill
\begin{Presented}
NuPhys2023, Prospects in Neutrino Physics\\
King's College, London, UK,\\ December 18--20, 2023
\end{Presented}
\vfill
\end{titlepage}
\def\thefootnote{\fnsymbol{footnote}}
\setcounter{footnote}{0}

\section{Introduction}

Super-Kamiokande (SK) is a 50 ktonne water Cherenkov detector located in the Kamioka mine in Japan. Neutrino interactions in the water produce relativistic charged particles, which emit Cherenkov light. These Cherenkov light signals are detected by 11,146 20-inch inward-facing photomultiplier tubes (PMTs) which instrument the inner detector volume. The inner detector is surrounded by a two-metre wide outer detector volume instrumented with around 1,885 8-inch PMTs. The active outer detector, in combination with the location of the detector 1000 km underground, serve to mitigate backgrounds from cosmic-ray muons created in the Earth's upper atmosphere.

There is an impressive history of discoveries at Kamioka, beginning in 1987 with the detection of neutrinos from a supernova explosion in the Kamiokande detector~\cite{Hirata1988}. The Super-Kamiokande experiment began running in April 1996 and, since 2020, has been running with gadolinium (Gd) dissolved in the pure water~\cite{Abe2021}. In 2020, 13 tonnes of gadolinium sulphate octahydrate were added to the detector, to achieve a concentration of 0.01\% Gd by mass (the SK VI phase). This marked the beginning of Super-Kamiokande Gadolinium (SK-Gd). In 2022, a further 26 tonnes of gadolinium sulphate octahydrate were added, to raise the concentration to 0.03\% Gd by mass (the SK VII phase).

Gd loading makes it possible to tag the neutrons produced in the inverse beta decay (IBD) interaction:
$$\rm \overline{\nu}_e + p \longrightarrow e^+ + n.$$

In pure water, the IBD neutron captures on hydrogen and this emits a 2.2~MeV gamma. The low light yield makes the neutron capture on hydrogen difficult to observe. In Gd-doped water, the neutron captures preferentially on gadolinium at concentrations greater than 0.01\%, and this increases the light yield from the capture of the IBD neutron by a factor of 3 to 4. In addition, the coincidence of the neutron capture with the positron signal is closer in time than in pure water, which enhances background rejection.

Inverse beta decay is the principal reaction of the antineutrino at the low energies of supernova neutrinos. The addition of Gd in Super-Kamiokande can improve pointing accuracy for a galactic supernova and increase the possibility of seeing pre-supernova neutrinos from Si-burning, and the diffuse supernova neutrino background (DSNB) from all supernovae since the beginning of time.

\section{Gadolinium for low-energy neutrino detection}

In a Cherenkov detector, positrons from the IBD interaction with a total energy above the Cherenkov threshold of $\sim$0.8~MeV emit a prompt signal. The neutrons from the IBD thermalize and are captured on nuclei in the medium, emitting a delayed signal. In pure water, the IBD neutrons capture on hydrogen, resulting in the delayed emission of a single 2.2~MeV gamma as the resulting deuteron decays to ground state. This occurs within a mean time of $\sim$200~$\rm \mu s$ of the prompt signal.

In Gd-doped water, the IBD neutron captures preferentially onto the Gd due to the very high neutron-capture cross section of Gd ($\sim$49,000~b) compared to hydrogen ($\sim$0.3~b). At a concentration of 0.1\% Gd ions, which can be achieved with the addition of 0.2\% gadolinium sulphate, 90\% of the neutrons may capture onto Gd. The remaining neutrons capture onto the hydrogen or sulphate. The subsequent decay of the Gd to ground state releases a cascade of gammas totalling $\sim$8~MeV in energy. These further interact in the water to produce Cherenkov light and the neutron-capture signal can be detected with a peak visible energy of around 4.5~MeV in a Gd-doped water Cherenkov detector, which is generally higher in energy than the positron signal. At a 0.1\% Gd concentration, the delayed neutron-capture signal occurs within a shorter mean time of $\rm \sim$30~$\rm \mu s$.

This combination of the prompt positron and higher-energy delayed neutron-capture signal within a short space and time results in a more easily detectable correlated signal in Gd-doped water compared to in pure water.

\begin{figure}
    \centering
    \includegraphics[width=0.9\linewidth]{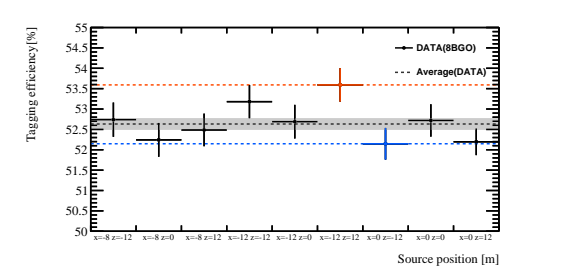}
    \vspace*{-13ex}  
    \begin{center}
    Preliminary
    \end{center}
    \vspace*{13ex}
    \caption{The measured neutron tagging efficiency in SK VII with a concentration of 0.03\% Gd using an Americium-Beryllium (AmBe) and bismuth germanate scintillator (BGO) crystals for a tagged neutron source.}
    \label{fig:efficiency}
\end{figure}

\begin{figure}
    \centering
    \includegraphics[width=0.9\linewidth]{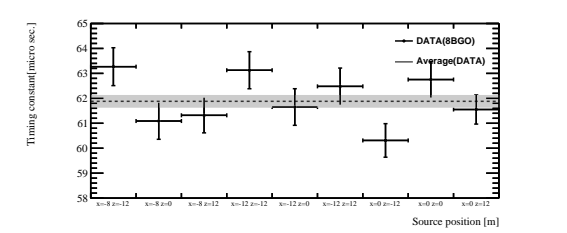}
    \vspace*{-13ex}  
    \begin{center}
    Preliminary
    \end{center}
    \vspace*{13ex}
    \caption{The measured time constant in SK VII with a concentration of 0.03\% Gd using an Americium-Beryllium (AmBe) and bismuth germanate scintillator (BGO) crystals for a tagged neutron source.}
    \label{fig:timeconstant}
\end{figure}

At the current concentration of 0.03\% Gd by mass, the measured neutron tagging efficiency is 53\%~(Fig.~\ref{fig:efficiency}) and the time between the positron and neutron capture signals is 62~$\mu$s~(Fig.~\ref{fig:timeconstant}).

\section{Core-collapse supernova neutrinos}

Core-collapse supernovae (CCSN) produce a neutrino burst which we can detect on Earth. Neutrinos from supernova 1987A in the Large Magellanic Cloud (50kpc) were detected in Kamiokande-II (11 neutrinos)\cite{Hirata1988}, IMB-3 (8 neutrinos)\cite{Haines1988} and Baksan (5 neutrinos)\cite{Alekseev1988}. The visible light came later. 

Neutrinos are produced at multiple stages in the evolution of the supernova: during initial matter infall and the subsequent shock revivial, and in the remnant proto-neutron star cooling phase. Neutrinos carry away more than 99\% of the energy from supernova explosions and the neutrino signal is produced a few minutes to several hours before the stellar explosion. This can serve to give advance warning to the wider supernova community.

A great deal of progress has been made recently in modelling the supernova explosion. Recent multi-dimensional supernova simulations successfully reproduce the CCSN explosion and several contributions to the explosion mechanism have been identified. There remains, however, a large disparity between CCSN models and there are many challenges, including understanding neutrino oscillations and the MSW effect in extreme high density.


Nevertheless, all models agree that neutrinos are key in the explosion mechanism and SK searches focus on three types of CCSN neutrino: pre-supernova neutrinos, supernova-burst neutrinos and the DSNB.

\section{Pre-supernova neutrinos - supernova early warning}

Neutrino emission increases as a massive star approaches the CCSN. During the silicon-burning phase, which precedes the CCSN and lasts for around a day, positron-electron annihilation is the dominant means of neutrino production in a pre-supernova (pre-SN) star, with weak processes such as beta decay increasing as the CCSN approaches. The silicon-burning phase is expected to produce electron neutrinos and electron antineutrinos up to around 4.7 MeV in energy, with an average energy of around 2~MeV~\cite{Odrzywolek2004}, depending on the model. The detection of these very low energy neutrinos relies on Gd to tag the neutrons from IBD interactions of pre-SN electron antineutrinos.

Detecting pre-SN neutrinos would give early warning of a supernova explosion and would relate pre-SN conditions in the interior of the star to subsequent explosion dynamics.

\begin{figure}
    \centering
    \includegraphics[width=0.7\linewidth]{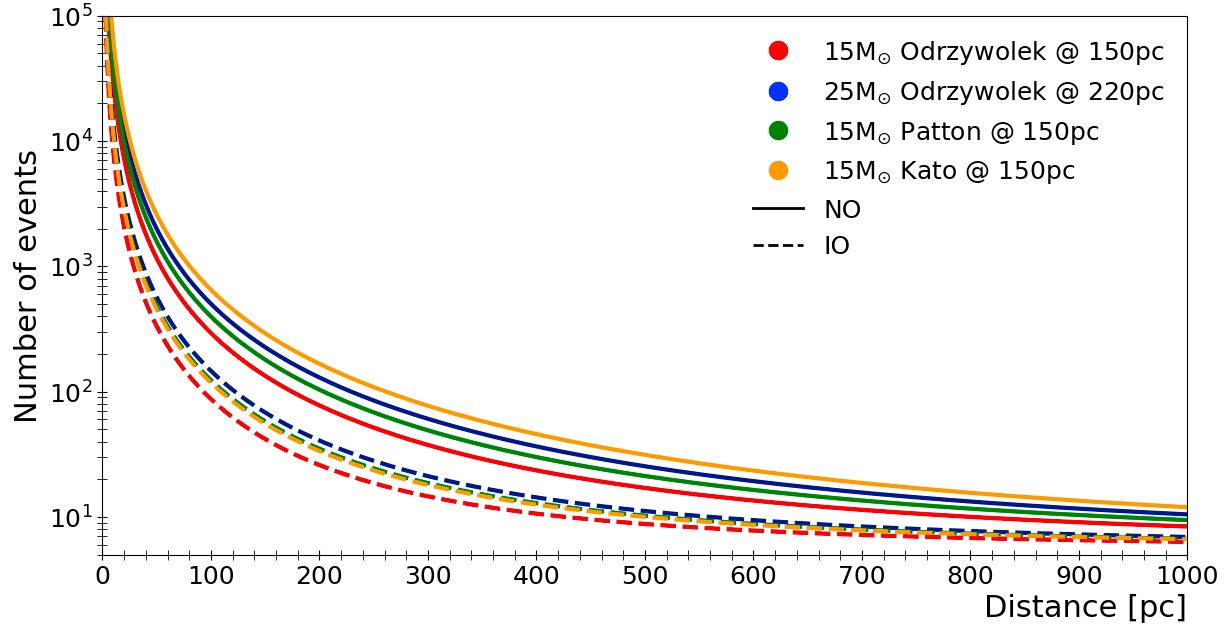}
    \vspace*{-30ex}  
    \begin{center}
    Preliminary
    \end{center}
    \vspace*{30ex}
    \caption{Expected number of events in Super-Kamiokande with 0.03\% Gd concentration in the 8 hours preceding a supernova for a number of models. See \cite{Machado2022} for references to the models.}
    \label{fig:preSN_distance}
\end{figure}

\begin{figure}
    \centering
    \includegraphics[width=0.7\linewidth]{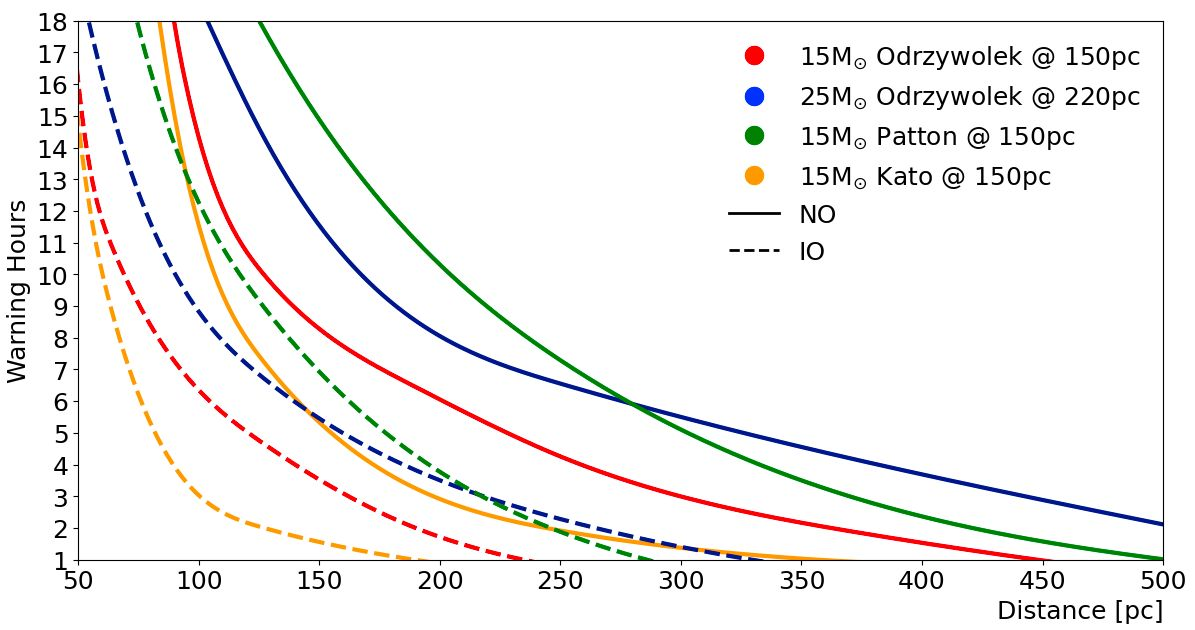}
    \vspace*{-30ex}  
    \begin{center}
    Preliminary
    \end{center}
    \vspace*{30ex}
    \caption{Expected number of hours warning at the 3~$\sigma$ level in Super-Kamiokande with 0.03\% Gd concentration for a number of models. See \cite{Machado2022} for references to the models.}
    \label{fig:preSN_warning}
\end{figure}

An online Super-Kamiokande pre-SN alert system went live in October 20221 and Super-Kamiokande also has a combined public pre-SN alert system with KamLAND. The SK pre-SN alert system is described in detail in \cite{Machado2022} In SK VII with a concentration of 0.03\% Gd, a pre-SN warning at the level of 3 $\sigma$ could be issued for a supernova up to 500 pc away with the most optimistic model~(Fig.~\ref{fig:preSN_distance}). Betelgeuse, at a distance of 150 pc, is a red supergiant star which is showing signs of nearing the end of its life. In the event of the CCSN explosion of Betelgeuse, the most optimistic model predicts that the pre-SN alert system could give up to 15 hours warning in advance of the CCSN explosion~(Fig.~\ref{fig:preSN_warning}).

\section{Supernova burst neutrinos - directional pointing}

Super-Kamiokande is predicted to be able to detect a burst of neutrinos from a CCSN explosion greater than 100~kpc away, depending on the CCSN model. Most of these events would be from IBD. In the event of a supernova, as much data as possible would need to be collected and stored for offline analysis. This data would provide input into developing supernova theory and refining the current models. In addition, a fast alarm is needed so that the supernova community can be alerted, within minutes, that the CCSN is imminent. This information would be enhanced by directional information in order to point in the direction of the supernova, within a few degrees of accuracy. For a Wolf Rayet star, there are just a few minutes between the neutrino and electromagnetic signals, so speed is key in evaluating the direction and issuing the alert.

\begin{figure}
    \centering
    \includegraphics[width=0.8\linewidth]{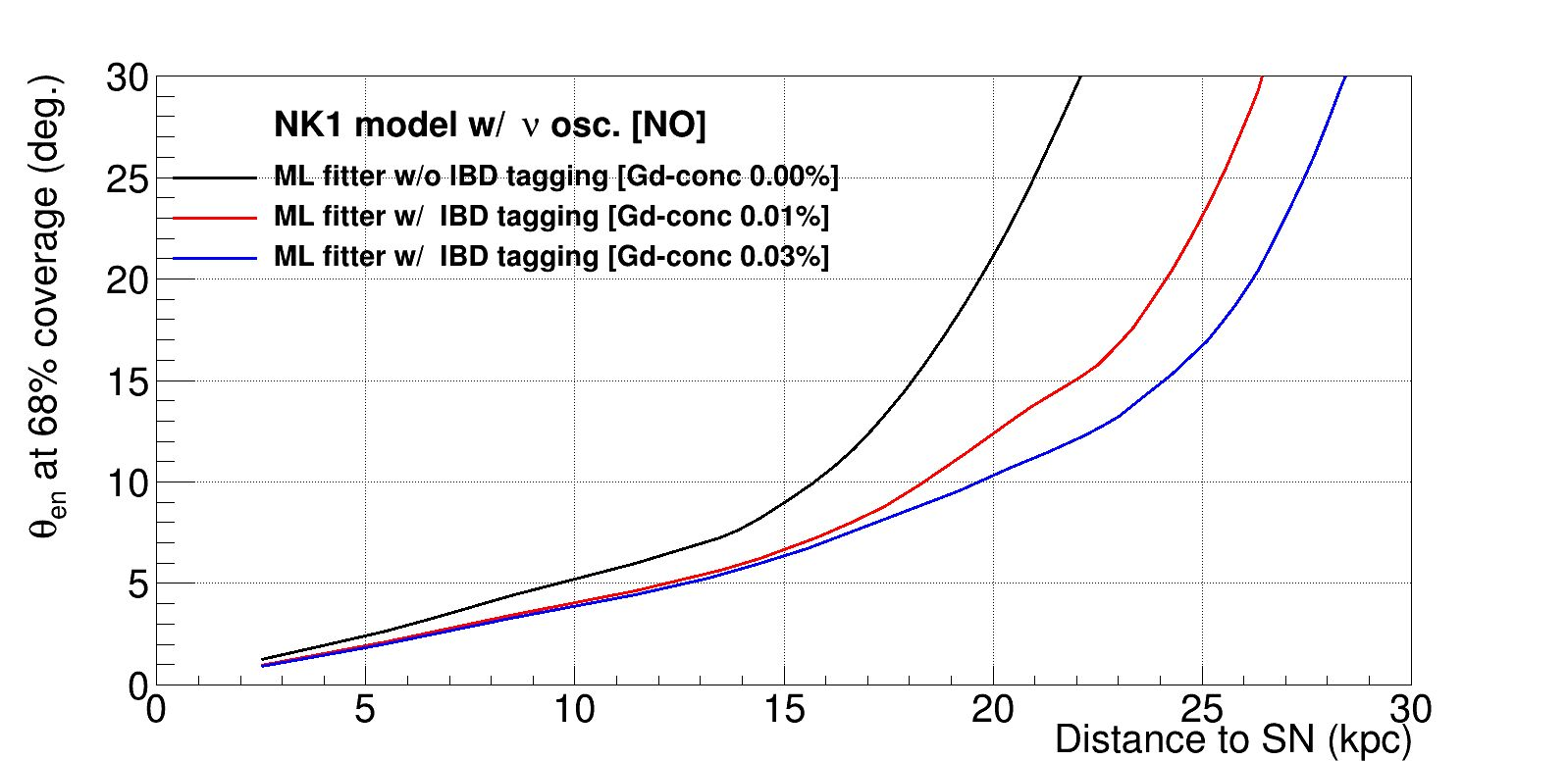}
    \vspace*{-10ex}  
    \begin{center}
    Preliminary
    \end{center}
    \vspace*{10ex}
    \caption{Pointing accuracy to a CCSN at 10 kpc using the NK1 model with neutrino oscillations under the assumtption of normal ordering.}
    \label{fig:pointing}
\end{figure}

Directional information comes from the elastic scattering of all flavours of supernova neutrinos on electrons:
$$\nu + e^- \longrightarrow \nu + e^-$$
In the elastic scattering interaction, the electron is largely forward-scattered in the direction of the incoming neutrino. As such, it is possible to reconstruct the direction of the CCSN from the electron directions. To do this, however, it is vital to isolate the elastic scattering interactions from the more numerous IBD events, which requires Gd in the water to tag the neutrons from the IBD events and thus isolate the elastic scatters, which do not produce neutrons. Pointing accuracy has been shown to increase with increasing Gd concentration for 0\%, 0.01\% (SK VI) and 0.03\% (SK VII) concentrations~(Fig.~\ref{fig:pointing}).

Super-Kamiokande has an online supernova burst alert system described in detail in \cite{Abe2016}. Data undergoes fast vertex, direction and energy reconstruction followed by IBD interaction selection. Processed data is then passed into the \textit{SNWatch} search for clusters of events in time, which are distributed uniformly across the detector as expected in the event of a CCSN explosion. Different levels of supernova alarm may be issued, depending on the measure of uniformity in SK and the number of events clustered in time. Preliminary results estimate that an alarm could be released in about a minute following a CCSN burst at 10 kpc and a pointing accuracy of three to four degrees could be achieved for the same CCSN with 0.03\% Gd concentration.

\section{Diffuse supernova neutrino background}

The DSNB consists of neutrinos from all CCSN since the beginning of time. Detecting the DSNB could give insight into the star formation rate, the CCSN rate as a function of redshift, average energy spectrum of CCSN burst neutrinos, average temperature inside the CCSN, and about black hole formation and the dim supernova rate.

DSNB models fold in the effects of different CCSN models with models of star formation rates, CCSN rates, black hole formation rates and dim supernova rates. As such there are many DSNB models to discriminate between.

\begin{figure}
    \centering
    \includegraphics[width=0.5\linewidth]{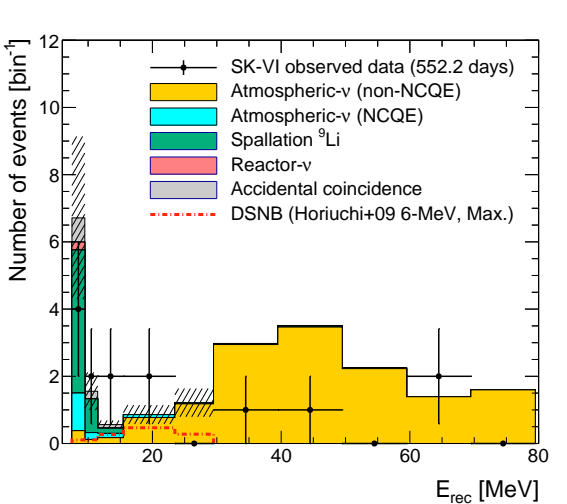}
    \vspace*{-20ex}  
    \begin{center}
    Preliminary
    \end{center}
    \vspace*{20ex}
    \caption{Signal and background remaining after spectrum-independent DSNB analysis in SK VI with 0.01\% Gd concentration, as a function of reconstructed energy.}
    \label{fig:DSNBevents}
\end{figure}

\begin{figure}
    \centering
    \includegraphics[width=0.5\linewidth]{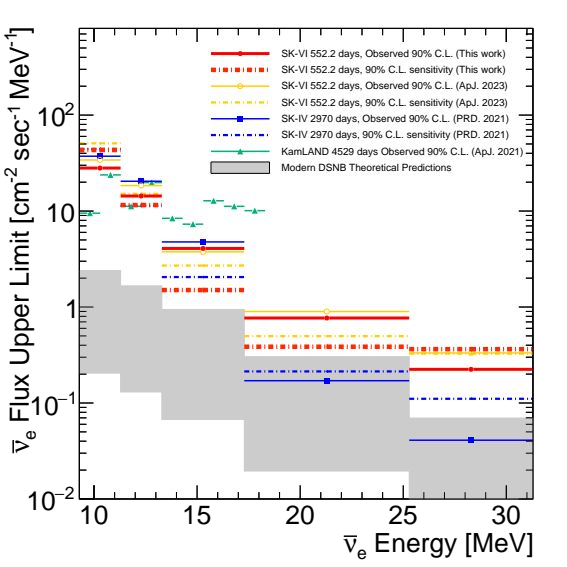}
    \vspace*{-30ex}  
    \begin{center}
    ~~~~~~~~~~~~~~~Preliminary
    \end{center}
    \vspace*{30ex}
    \caption{DSNB flux upper limits for different SK run periods, as a function of electron antineutrino energy.}
    \label{fig:DSNB_spectrum}
\end{figure}

The window for observing the DSNB is narrow due to backgrounds from solar, reactor and atmospheric neutrinos as well as non-neutrino cosmogenic muon-induced backgrounds and backgrounds due to radioactive decay. As such it is necessary to aggressively drive down backgrounds. The latest results for SK VI with 0.01\% Gd concentration, with improved background rejection compared to~\cite{Harada2023}, found 14 DSNB events~(Fig.~\ref{fig:DSNBevents}) and set a new limit at below 17 MeV in just a fifth of the observation time in the pre-Gd phase~(Fig.~\ref{fig:DSNB_spectrum}). Improvements in background rejection included a neural network-based neutron tagging method, and an updated atmospheric neutrino simulation and rejection based on the \textit{multiple scattering goodness} variable as calculated in \cite{Abe2016b}.

\section{Conclusions and Outlook}
Super-Kamiokande Gd is a new era in CCSN detection which has already led to new results in supernova neutrino detection:
\begin{itemize}
    \item Potential for detection of neutrinos from pre-supernova stars up to 800 pc away at 3$\sigma$, and from
Betelgeuse up to 15 hours before the explosion.
    \item Supernova pointing accuracy has been improved to a few degrees for a 10-kpc supernova.
    \item A supernova alert could be issued in less than a minute for a supernova burst at 10 kpc.
    \item New limits on DSNB at $<$~17 MeV in SK VI (0.01\% Gd) in a fifth of the observation time
compared to the pre-Gd phase.
\end{itemize}

Work is now focused on using the increased Gd concentration in SK VII (0.03\%), all
the time innovating and improving the existing analyses.

\begingroup\raggedright\endgroup

\end{document}